\documentclass[onecollarge]{article}
\usepackage{latexsym}
\usepackage{amsfonts,amsbsy,amssymb,epsfig}
\usepackage{graphics,graphicx}
\usepackage{amsmath}
\usepackage{graphicx}
\usepackage{color}

\begin{document}

\title{\bf Primeval symmetries}
\author{\textbf{R. Aldrovandi},$^{1}$ \textbf{R. R. Cuzinatto},$^{1, }$\thanks{rodrigo@ift.unesp.br}
\, and \textbf{L. G. Medeiros}$^{1}$
\bigskip \\$^{1}${\small Instituto de
F\'{\i}sica Te\'{o}rica, Universidade Estadual Paulista}\\{\small
Rua Pamplona 145, CEP 01405-900, S\~{a}o Paulo, SP,
Brazil.\bigskip}}
\date{}
\maketitle

\begin{abstract}
\noindent {A detailed examination of the Killing equations in
Robertson--Walker coordinates shows how the addition of matter
and/or radiation to a de Sitter Universe breaks the symmetry
generated by four of its Killing fields. The product $U = a^2
\,{\dot H}$ of the squared scale parameter by the time-derivative of
the Hubble function encapsulates the relationship between the two
cases: the symmetry is maximal when $U $ is a constant, and reduces
to the 6-parameter symmetry of a generic
Friedmann--Robertson--Walker model when it is not. As the fields
physical interpretation is not clear in these coordinates,
comparison is made with the Killing fields in static coordinates,
whose interpretation is made clearer by their direct relationship to
the Poincar\'e group generators via Wigner-In\"on\"u contractions.}

\end{abstract}

\vskip 0.5cm


\section{Introduction}


Symmetry considerations are enough to fix the general form
\begin{equation}
ds^{2} = c^{2} dt^{2} - a^{2}(t) \left[ \frac{dr^{2}}{1 - \kappa r^{2}} +
r^{2} d\theta^{2} + r^{2} \sin^{2}\theta d \phi^{2} \right]  \label{ds2 FRW}
\end{equation}
of the Friedmann--Robertson--Walker (FRW) interval~\cite{Nar93,Wei72}.
Homogeneity and isotropy of the space--section ensure a minimum of 6
symmetry generators, but this number can be higher for particular
expressions of the scale parameter $a(t)$. The expression above holds for
any spacetime whose space--section is homogeneous and isotropic and
includes, consequently, de Sitter ($dS$) spacetimes. Some coordinate
systems  may be specifically more convenient for the latter, such as the
static line element~ \cite{To87}
\begin{equation}
ds^{2} = {\textstyle{\left( 1 - \frac{\bar{r}^2}{L^2} \right) }} c^{2} d\bar{%
t}^{2} - \frac{d\bar{r}^{2}}{ \left( 1 - \frac{\bar{r}^2}{L^2} \right) } -
\bar{r}^{2} d\theta^{2} - \bar{r}^{2} \sin^{2}\theta d \phi^{2} ,
\label{ds2 dS}
\end{equation}
but expression (\ref{ds2 FRW}) provides a simpler, unified view of
all the homogeneous isotropic Universes. Our objective here will be
to analyse the isometries in the general case, with emphasis on the
relationship between the de Sitter ten-parameter group of motions
and the FRW six-parameter symmetry.

We shall adopt a working picture which has formal~---~and, nowadays,
observational~---~advantages: the Universe starts in an inflationary era
modeled by a pure de Sitter model with cosmological constant $\Lambda$,
whose maximal symmetry is later partially broken by the addition of
radiation and matter. It is subsequently described by the standard FRW model
with a cosmological constant.

The ten-generator de Sitter group is a deformation~\cite{gursey,gil}
of the Poincar\'e group ${\mathcal{P}}$, and its Lie algebra has, in
consequence, a preferred basis in which the generators have clear
physical interpretations. The Poincar\'e group~---~the group of
motions~\cite{Eis49} of flat Minkowski spacetime $M$~---~is an
In\"on\"u-Wigner contraction~\cite{In62} of the de Sitter
group~---~the group of motions of de Sitter spacetime $dS$. And
${\mathcal{P}} = {\mathcal{L}} \oslash {\mathcal{T}}$, the
semi-direct product of the Lorentz  group ${\mathcal{L}} $ and the
group ${\mathcal{T}}$ of translations on $M$. Minkowski spacetime is
the quotient $M = {\mathcal{P}}/{\mathcal{L}}$. The
 ${\mathcal{P}}$ Lie algebra has a general form given by
commutators
\begin{gather}
[{\mathcal{L}}, {\mathcal{L}}] = {\mathcal{L}} \, ; \quad
[{\mathcal{L}}, {\mathcal{T}}] = {\mathcal{T}} \, ; \quad
[{\mathcal{T}}, {\mathcal{T}}] = 0\, ,  \notag
\end{gather}
where ${\mathcal{L}}$ and ${\mathcal{T}}$ represent generic generators of the Lorentz sub-group
 of the translation sub-group.

Now, each $dS$ spacetime is characterized by a pseudo-radius (or
horizon) $L$. The de Sitter group has also a Lorentz sub-group, but
the translations differ from their Poincar\'e counterparts: their
generators are dimensionless and do not constitute a sub-group. The
general form of the Lie algebra is
\begin{gather}
[{\mathcal{L}}, {\mathcal{L}}] = {\mathcal{L}} \, ; \quad
[{\mathcal{L}}, {\mathcal{T}}] = {\mathcal{T}}  \, ; \quad
[{\mathcal{T}}, {\mathcal{T}}] = {\textstyle{\frac{1}{L^2}}}\,
{\mathcal{L}}\, .  \notag
\end{gather}
The In\"on\"u-Wigner contraction corresponds to taking the limit $L \to
\infty$, after changing the ${\mathcal{T}}$ generators' dimensions through
multiplication by convenient constants. Comparison of the two commutation
tables provide clear physical meanings for the $dS$ generators from those,
well-known, of the ${\mathcal{P}}$ generators. That is, for example, where
viewing the ${\mathcal{T}}$ generators as translations on $dS$ space comes
from.

In   $FRW$ coordinates the homogeneous
cosmological models are unified and more easily compared with each other~\cite%
{CE97,ACM05}, but  the $dS$ generators  have not  clear
interpretations. The Killing fields for metric (\ref{ds2 dS}) have an immediate relationship to
${\mathcal{P}}$ generators~---~precisely through the In\"on\"u-Wigner contraction.

Our aim here is not only to exhibit the Killing fields for metric (\ref{ds2 FRW}),
using the FRW coordinates to correlate
 the $dS$ and general $FRW$ symmetries: it is also to
give them simple physical interpretations. With that in mind we shall
also recall the Killing fields for metric (\ref{ds2 dS}), establish the
transformations between the two systems, and examine the
contractions in both cases.

The Killing fields in FRW coordinates are given in Section \ref{sec-KillFRW}.
It turns out that the vanishing or not of their timelike
components tally with the constancy or not
of the parameter $U(t)=a^{2}(t)\dot{H}$. There are ten Killing fields
when $U$\ is time-independent, corresponding to the
$dS$ case; otherwise that number  drops to six,
corresponding to the FRW models with a cosmological
constant ($\Lambda $-FRW Universes).  Details are given in  Appendix
\ref{app-Killcomponents}. Section \ref%
{sec-KilldS} discusses  the static coordinates Killing fields. The
transformations relating both
coordinate systems are
given in Appendix \ref{app-GRT}. Section \ref{sec-Contractions} is
devoted to the contractions of the generators and their physical interpretation.


\section{FRW and dS symmetries in comoving coordinates}

\label{sec-KillFRW}

A Killing vector field ${\boldsymbol{\xi }^{(P)}}$ with $P=1,2,...,n$, $n$
being the number of parameters of the symmetry group, will be here indicated
by its covariant components $\xi _{\alpha }^{(P)}=\xi _{\alpha
}^{(P)}(x^{0},x^{1},x^{2},x^{3})$, i.e., ${\boldsymbol{\xi }^{(P)}}=(\xi
_{0}^{(P)},\xi _{1}^{(P)},\xi _{2}^{(P)},\xi _{3}^{(P)})$, in terms of which
the Killing equations take on the simple forms
\begin{equation}
\nabla _{\alpha }\xi _{\beta }+\nabla _{\beta }\xi _{\alpha }=\partial
_{\alpha }\xi _{\beta }^{(P)}+\partial _{\beta }\xi _{\alpha
}^{(P)}-2\,\Gamma ^{\gamma }{}_{\alpha \beta }\,\xi _{\gamma }^{(P)}=0\,
\label{Killing eqs} .
\end{equation}%
 Each Killing
field~---~one for each independent integration constant~---~will be a vector field~\cite{DNF79,AP95,KN63}
\begin{equation}
{\boldsymbol{\xi }^{(P)}}=\xi ^{(P)\alpha }\partial _{\alpha } \, .
\label{Killing fields}
\end{equation}%
 We shall  from now on omit index ${}^{(P)}$,
take $c=1$  and, in Eq.(\ref{ds2 FRW}), adopt the usual practice~%
\cite{LL75} of considering also coordinate $x^{1}=r$ as non-dimensional,
besides $x^{2}=\theta $\ and $x^{3}=\phi $. The length/time dimension will
be carried by the scale function $a\left( t\right) $. Equations (\ref%
{Killing eqs}) turn out to be
\begin{align}
&\partial _{0}\xi _{0}=0~;  \label{I} \\
&\partial _{1}\xi _{0}+\partial _{0}\xi _{1}-2H~\xi _{1}=0~;  \label{II} \\
&\partial _{2}\xi _{0}+\partial _{0}\xi _{2}-2H~\xi _{2}=0~;  \label{III} \\
&\partial _{3}\xi _{0}+\partial _{0}\xi _{3}-2H~\xi _{3}=0~;  \label{IV} \\
&\partial _{2}\xi _{1}+\partial _{1}\xi
_{2}-{\textstyle{\frac{2}{r}}}\,\xi
_{2}=0~;  \label{V} \\
&\partial _{3}\xi _{1}+\partial _{1}\xi
_{3}-{\textstyle{\frac{2}{r}}}\,\xi
_{3}=0~;  \label{VI} \\
&\partial _{3}\xi _{2}+\partial _{2}\xi _{3}-2\cot \theta ~\xi
_{3}=0~;
\label{VII} \\
&\left( 1-\kappa r^{2}\right) ~\partial _{1}\xi _{1}-\kappa r~\xi
_{1}-a^{2}H~\xi _{0}=0~;  \label{VIII} \\
&\partial _{2}\xi _{2}+r~\left( 1-\kappa r^{2}\right) ~\xi
_{1}-a^{2}H~r^{2}~\xi _{0}=0~;  \label{IX} \\
&\partial _{3}\xi _{3}+\sin \theta \cos \theta ~\xi _{2}+r~\left(
1-\kappa r^{2}\right) \sin ^{2}\theta ~\xi _{1}-a^{2}H~r^{2}\sin
^{2}\theta ~\xi _{0}=0~.  \label{X}
\end{align}

The fastidious process to find the analytic solutions follows the standard
procedure of taking crossed derivatives of the equations and eliminating
unwanted terms. We shall only comment on a particular case: take $%
\partial _{0}$\ of (\ref{VIII}), use (\ref{I}) in order to eliminate $%
\partial _{0}\xi _{0}$, and (\ref{II}) to substitute $\partial _{0}\xi _{1}$%
. This leads, after using (\ref{VIII}) again to replace $\partial _{1}\xi
_{1}$, to the second-order equation for the $r$-dependence of  component $%
\xi _{0}$:
\begin{equation}
(1-\kappa r^{2})\,\partial _{1}\,\partial _{1}\,\xi _{0}-\kappa
\,r\,\partial _{1}\,\xi _{0}+U(t)\,\xi _{0}=0\,,  \label{diff eq xi0 r}
\end{equation}%
where we have introduced the symmetry-controling parameter
\begin{equation}
U\left( t\right) =a^{2}\dot{H}~,  \label{U}
\end{equation}%
with $H\left( t\right) =\frac{\dot{a}\left( t\right) }{a\left( t\right) } $ the usual Hubble function. The Friedmann equations are
\begin{eqnarray}
H^{2} &=&\frac{8\pi G}{3}\rho +\frac{\Lambda }{3}-\frac{\kappa }{a^{2}}~,
\label{Fried H} \\
U \equiv a^{2}\dot{H} &=&-\frac{3}{2}\left[ \frac{8\pi G}{3}\left( \rho +p\right) %
\right] a^{2}+\kappa ~,  \label{Fried H dot}
\end{eqnarray}%
where $\Lambda$ is  the  cosmological constant ($=\frac{3}{L^{2}}$ in terms
of the de Sitter radius $L$),   $p$ and  $\rho $ are the source pressure and density, and $\kappa $ is the curvature parameter  with  values in the set $\left\{ -1,0,+1\right\} $.

The role of function $U\left( t\right) $\  becomes clear from
taking $\partial _{0}$\ of (\ref{diff eq xi0 r}%
) and using (\ref{I}) to obtain
\begin{equation}
\dot{U}~\xi _{0}=0~.  \label{constraint}
\end{equation}%
Hence, either $\xi _{0}=0$ or $\dot{U}=0$, or both. For the
time being, let us focus on the second case, when $U$ is a constant.
 Equation (\ref{Fried H dot}) says that this will happen when $\left( \rho +p\right) = 0$,
so that $U=\kappa $. Requirement $\left( \rho +p\right) =0$ is fulfilled
when \textbf{(i)} $\rho =p=0$,   the vacum solution of (\ref{Fried H}, %
\ref{Fried H dot}), by definition the dS model;  \textbf{(ii)} $%
p=-\rho $, an exotic equation of state which just simulates the cosmological constant.
Whichever the choice, $a$\ and $H$\ will be the scale factor and the Hubble
function for the dS solution \cite{ACM05}:%
\begin{equation}
a\left( t\right) =A\cosh \frac{t}{L}+\sqrt{A^{2}-\kappa L^{2}}\sinh \frac{t}{%
L}~,  \label{a dS}
\end{equation}%
with the arbitrary initial condition $A=a\left( 0\right) $ and%
\begin{equation}
H\left( t\right) =\frac{1}{L}-\frac{2\kappa L}{\kappa L^{2}+e^{2\frac{t}{L}%
}\left( A+\sqrt{A^{2}-\kappa L^{2}}\right) ^{2}}~.  \label{H dS}
\end{equation}%

We see thus that $\xi _{0}\neq 0$ is precisely the case of de Sitter
solutions. Otherwise, $a\left( t\right) $\ is unconstrained and may be any
of the solutions for the Standard Model. Summing up:
\begin{equation*}
\xi _{0}\quad
\begin{cases}
=0\qquad {\mbox{when $U$ is not a constant (FRW spacetime)}} \\
\neq 0\qquad {\mbox{when $U = \kappa$ is a constant  (dS spacetime)}}\,.%
\end{cases}%
\end{equation*}%
The particular expressions of the scale parameter $a(t)$ for which the
symmetry is higher are now clear: they are those for which $U=(a\,{\ddot{a}}-%
{\dot{a}}^{2})$ is an integral of motion.

The detailed procedure leading to the component $\xi _{0}$ is the subject of
Appendix \ref{app-Killcomponents}. The other components $\xi _{i}$\ are
obtained by analogous techniques, and are simply listed in the same
appendix, where it is also indicated how to calculate the Killing fields in their final forms, namely%
\begin{eqnarray}
S_{0} &=&\sqrt{1-\kappa r^{2}}\left( \partial _{t}-H~D\right)  \notag \\
S_{i} &=&x^{i}~\left( \partial _{t}-H~D\right) +\kappa ~H~\partial _{i}-{%
\textstyle{\ \frac{\left( 1-\kappa ^{2}\right) }{2}\left( \frac{1}{a^{2}H}%
-H~r^{2}\right) }}\partial _{i}  \notag \\
T_{i} &=&-~\sqrt{1-\kappa r^{2}}~\partial _{i}  \label{generators FRW} \\
J_{i} &=&\epsilon _{ij}^{\;\;k}x^{j}\partial _{k}~.  \notag
\end{eqnarray}%
The Hubble function $H$ will be given by (\ref{H dS}) whenever it turns up:
it appears  just in  $S_{0}$\ and $S_{i}$, present only in the dS case.
Some interpretation is possible at this stage. Each field $T_{i}$ is a translation generator, the factor $\sqrt{1-\kappa r^{2}}$
accounting for curvature. The same would hold for $S_{0}$, a
time-translation on a curved space from which space expansion (term $H~D$)
is subtracted. The $J_{i}$'s have the usual  aspect of rotation generators.
Generators $S_{i}$, however, elude any simple putting into words.

The commutation table for the fields (\ref{generators FRW}) exhibits
the dS group algebra in comoving coordinates:
\begin{eqnarray}
\left[ S_{0},S_{i}\right] &=&-~\kappa ~{\textstyle{\frac{1}{L^{2}}}}%
~T_{i}-\left( 1-\kappa ^{2}\right) {\textstyle{\frac{1}{L}}}~S_{i}
\label{[S0,Si]} \\
\left[ S_{0},T_{i}\right] &=&-~\kappa ~S_{i}+\left( 1-\kappa ^{2}\right) {%
\textstyle{\frac{1}{L}}}~T_{i}  \label{[S0,Ti]} \\
\left[ S_{0},J_{i}\right] &=&0  \label{[S0,Ji]} \\
\left[ S_{i},S_{j}\right] &=&\kappa ~{\textstyle{\frac{1}{L^{2}}}}~\epsilon
_{ij}^{\;\;k}~J_{k}  \label{[Si,Sj]} \\
\left[ S_{i},T_{j}\right] &=&-~\delta _{ij}~S_{0}-\left( 1-\kappa
^{2}\right) {\textstyle{\frac{1}{L}}}~\epsilon _{ij}^{\;\;k}~J_{k}
\label{[Si,Tj]} \\
\left[ S_{i},J_{j}\right] &=&-~\epsilon _{ij}^{\;\;k}~S_{k}
\label{[Si,Jj]} \\
\left[ T_{i},T_{j}\right] &=&-~\kappa
~\epsilon_{ij}^{\;\;k}~J_{k}
\label{[Ti,Tj]} \\
\left[ T_{i},J_{j}\right] &=&-~\epsilon _{ij}^{\;\;k}~T_{k}  \label{[Ti,Jj]} \\
\left[ J_{i},J_{j}\right] &=&\epsilon _{ij}^{\;\;k}~J_{k}~.  \label{[Ji,Jj]}
\end{eqnarray}

At first sight, this table has a plain enough meaning. Equations (\ref%
{[S0,Si]}-\ref{[Si,Jj]}) involve those generators which are   absent in
the generic FRW case. Only Eqs.(\ref{[Ti,Tj]}-\ref{[Ji,Jj]}) appear in
that case, and represent the groups of motions of a sphere $S^3$ (if  $%
\kappa = +1$), of a hyperbolic 3-space (if $\kappa = - 1$) and of the
euclidian space ${\mathbb{E}}^3$ (if  $\kappa = 0$)~---~just the expected
Universe space-sections in the Standard Model~\cite{Ga94}.
These groups are, respectively, $SO(4) = SO(3) \otimes SO(3)$, $SO(3,1) = {%
\mathcal{L}}$ and the  Euclidean Group. They also exhibit the
bundle aspect~\cite{AP95} of the spaces involved, important for its
generality: the translations are ``horizontal'', responsible for the space
homogeneity. They span the space itself, and their commutator (\ref{[Ti,Tj]}%
) exhibits the space curvature $\kappa$ as a tag multiplying the
``vertical'', rotation-related generators.

And here comes a difficulty: the same holds for de Sitter spaces. Their
translation commutators (\ref{[S0,Si]}) and (\ref{[Si,Sj]}) display the
spacetime curvature $1/L^2$ in their vertical components, along the
Lorentz generators. The trouble is that the translation de Sitter generators
$\{S_{k}\}$ are not the translation generators $\{T_{k}\}$ of the $\Lambda$%
-FRW Universe. This seems to preclude any possible passage from a $dS$ model
to a Standard Model. Notice, however, that any linear combination of Killing
fields is also a Killing field. It might happen that the de Sitter generator
basis provided by the FRW coordinates be not the most convenient, but some
combination which, in the passage, also reduces to those appearing in Eqs.(%
\ref{[Ti,Tj]}-\ref{[Ji,Jj]}). To examine this question, a better
understanding of the physical meanings of the above Killing fields is
necessary. We proceed then to obtain clearer
interpretations for them by studying their relationships with the Poincar\'e
generators. We will then: \textit{(i)} repeat the procedure for the dS
solution written in static coordinates; and \textit{(ii)} convert the $%
dS $ generators from the static coordinates to the FRW comoving coordinates,
consistently recovering $S_{0}$, $S_{i}$, $T_{i}$\ and $J_{i}$. Step \textit{%
(ii)} requires, however, a previous step \textit{(iii)}: to obtain the
general transformation between FRW coordinates~---~for the three possible
values of $\kappa $~---~and static dS invariant coordinates. Steps \textit{%
(i)} and \textit{(ii)} are undertaken in the following section though, for
economy, the results are just listed. The transformations themselves are
left to Appendix \ref{app-GRT}. Relations to the Poincar\'e generators
are discussed in the subsequent sections.


\section{Symmetry relations  in static and
comoving coordinates}
\label{sec-KilldS}


In FRW  coordinates the radial coordinate $r$\ is dimensionless, while $%
a\left( t\right) $ and $t$\ (with $c=1$) have length dimension. In the
static coordinates of line-element~(\ref{ds2 dS}), $\bar{r}$ has
length dimension, as well as the de Sitter horizon pseudo-radius $L$ and
time $\bar{t}$. The angular coordinates are the same in both systems.

Integration of the Killing equations for metric (\ref{ds2 dS}) leads to the
covariant components $\bar{\xi}_{\mu }$\ as functions of the coordinates $%
\bar{x}^{0}=\bar{t}$, $\bar{x}^{1}=\bar{r}$, $\bar{x}^{2}=\theta $\ and $%
\bar{x}^{3}=\phi $. The contravariant Killing
components $\bar{\xi}^{\mu }=\bar{g}^{\mu \nu }\bar{\xi}_{\nu }$ will then
give directly  the generators $\bar{X}_{N}=\bar{\xi}_{N}^{\;\;\;\mu }\bar{%
\partial}_{\mu }$\ associated to the ten integration constants. After a
transformation from the static $\left( \bar{r},\theta ,\phi \right) $ to
Cartesian coordinates (now with length dimension) $\bar{x}=\bar{r}%
\sin \theta \cos \phi $; $\bar{y}=\bar{r}\sin \theta \sin \phi $; $\bar{z}=%
\bar{r}\cos \theta $, the operators $\bar{X}_{N}$\ are obtained as:
\begin{align}
T_{0}& =\partial _{\bar{t}}~;  \notag \\
B_{i}& ={\textstyle-{\frac{1}{\sqrt{1-\frac{\bar{r}^{2}}{L^{2}}}}\cosh \frac{%
\bar{t}}{L}~\bar{x}^{i}~\partial _{\bar{t}} - L\sqrt{1-\frac{\bar{r}^{2}}{L^{2}%
}}\sinh \frac{\bar{t}}{L}~\bar{\partial}_{i}~;}}  \notag \\
P_{i}& ={\textstyle{\ -~\frac{1}{L}\frac{1}{\sqrt{1-\frac{\bar{r}^{2}}{L^{2}}%
}}\sinh \frac{\bar{t}}{L}~\bar{x}^{i}~\partial _{\bar{t}}-\sqrt{1-\frac{\bar{%
r}^{2}}{L^{2}}}\cosh \frac{\bar{t}}{L}~\bar{\partial}_{i}~;}}  \notag \\
\bar{J}_{i}& =\epsilon _{ij}^{\;\;k}\bar{x}^{j}\bar{\partial}_{k}~,
\label{geradores dS}
\end{align}%
The physical interpretation of these symmetries arises from the analysis of
the limit\ $L\rightarrow \infty $, when these dS operators should reduce to
the Poincar\'{e} generators: boosts, rotations and temporal and spatial
translations. The commutation table of the generators above is:
\begin{equation}
\begin{array}{lll}
\left[ T_{0},B_{i}\right] =-~P_{i} & \left[ T_{0},P_{i}\right] ={\textstyle{%
\ -\frac{1}{L^{2}}}~}B_{i}\,\quad \quad  & \left[ T_{0},\bar{J}_{i}\right] =0
\\
\left[ B_{i},B_{j}\right] =\epsilon _{ij}^{\;\;k}\bar{J}_{k} & \left[
B_{i},P_{j}\right] =-~\delta _{ij}~T_{0} & \left[ B_{i},\bar{J}_{j}\right]
=-~\epsilon _{ij}^{\;\;k}B_{k} \\
\left[ P_{i},P_{j}\right] =-~{\textstyle{\frac{1}{L^{2}}}}~\epsilon
_{ij}^{\;\;k}\bar{J}_{k}\,\quad \quad  & \left[ P_{i},\bar{J}_{j}\right]
=-~\epsilon _{ij}^{\;\;k}P_{k} & \left[ \bar{J}_{i},\bar{J}_{j}\right]
=\epsilon _{ij}^{\;\;k}\bar{J}_{k}~.%
\end{array}
\label{algebra dS}
\end{equation}

Both sets of operators $\left( S_{0},S_{i},T_{i},J_{i}\right) $ and $\left(
T_{0},B_{i},P_{i},\bar{J}_{i}\right) $ are generators of the dS group. The
coordinate transformations connecting both sets are also responsible for the
conversion of the FRW line element (\ref{ds2 FRW}) into the de Sitter static
interval (\ref{ds2 dS}). Appendix \ref{app-GRT} describes in detail how to
obtain these general transsformation laws, valid for all values of $\kappa $
and for an arbitrary initial condition $a\left( 0\right) =A$\ for the scale
factor. They are the \textit{generalized Robertson transformation},\footnote{%
$\tau(\kappa)$ is an integration constant. The convenience of a
complex number for the case $\kappa =+1$ is justified in Appendix
\ref{app-GRT}.}
\begin{eqnarray}
\bar{r} &=&r~a\left( t\right) ~,\;\;a\left( t\right) =A\cosh \frac{t}{L}+%
\sqrt{A^{2}-\kappa L^{2}}\sinh \frac{t}{L}~,  \notag \\
\bar{t} &=&L\ln {\textstyle{\ \left[ \tau \left( \kappa \right) \frac{%
a\left( t\right) \left( H\left( t\right) +\frac{\sqrt{1-\kappa r^{2}}}{L}%
\right) }{\sqrt{1-\frac{\left[ ra\left( t\right) \right] ^{2}}{L^{2}}}}%
\right] ~,\;\;}}\tau \left( \kappa \right) =\left\{
\begin{array}{l}
\frac{L}{2A}~,\;\;\text{\ }\kappa =0 \\
1~,\;\;\;\kappa =-1 \\
\frac{1}{i}~,\;\;\;\kappa =+1%
\end{array}%
\right. ~.  \label{TRG}
\end{eqnarray}

From (\ref{TRG}) we write dS Killing fields in comoving coordinates
$\left( S_{0},S_{i},T_{i},J_{i}\right) $ in terms of dS Killing
fields in static coordinates $\left(
T_{0},B_{i},P_{i},\bar{J}_{i}\right) $:\footnote{Acknowledgement is
due to an unknown referee for calling the authors' attention to
these general equalities.}
\begin{eqnarray}
S_{0} &=&T_{0}  \notag \\
S_{i} &=&{\textstyle{  \left( 1-\frac{\kappa ^{2}}{2}\right) \left[ \tau \left( \kappa
\right) -  \frac{ \kappa}{\tau \left( \kappa \right) }\right] }}\, P_{i} +
{\textstyle{  \left( 1-%
\frac{\kappa ^{2}}{2}\right) \left[ \tau \left( \kappa \right) +  \frac{%
\kappa}{\tau \left( \kappa \right) }\right] \frac{1}{L} }} \, B_{i}  \notag \\
T_{i} &=& {\textstyle{ \frac{1}{2}\left[ \frac{1}{\tau \left( \kappa \right) }+\kappa
~\tau \left( \kappa \right) \right] L }}\,  P_{i} - {\textstyle{  \frac{1}{2}\left[ \frac{1}{\tau
\left( \kappa \right) }-\kappa ~\tau \left( \kappa \right) \right] }}  B_{i}
\label{FRW to dS} \\
J_{i} &=&\bar{J}_{i}  \notag .
\end{eqnarray}%
Substituting $\tau \left( \kappa \right) $ as given in (\ref{TRG})
for each value of $\kappa $, we find the results of Table 1.


\begin{center}
\begin{tabular}{||c||}
\hline\hline
\textit{Correspondence between dS generators} \\
\hline\hline
\begin{tabular}{|c|c|}
\hline static$\text{ dS coordinates}$ & $\text{comoving FRW
coordinates}$ \\ \hline $T_{0}$ & $S_{0}$ \\ \hline $P_{i}$ &
\begin{tabular}{lc}
$\kappa =0$ & $\frac{A}{L}S_{i}+\frac{1}{2A}T_{i}$ \\
$\kappa =-1$ & $S_{i}$ \\
$\kappa =+1$ & $i~S_{i}$%
\end{tabular}
\\ \hline
$B_{i}$ &
\begin{tabular}{lc}
$\kappa =0$ & $A~S_{i}-\frac{L}{2A}T_{i}$ \\
$\kappa =-1$ & $-~T_{i}$ \\
$\kappa =+1$ & $i~T_{i}$%
\end{tabular}
\\ \hline
$\bar{J}_{i}$ & $J_{i}$ \\
\end{tabular}
\\ \hline\hline
\end{tabular}

\bigskip \emph{Table 1: Relations between dS generators in static (barred)
coordinates and FRW (unbarred) coordinates. }
\end{center}


\bigskip

The comoving operators $S_{0}$, $S_{i}$, $T_{i}$, and $J_{i}$\ are obtained,
directly or through linear combinations, from the dS generators in the
familiar static coordinates. They will tend, consequently, to well-defined
Poincar\'{e} generators when $L\rightarrow \infty $, just as $T_{0}$, $P_{i}$%
, $I_{i}$\ and $\bar{J}_{i}$. Next section is concerned with this point.


\section{Contraction from dS to Poincar\'{e} algebras}

\label{sec-Contractions}



\subsection{Limit $L\rightarrow \infty $\ in static coordinates}

\label{sec-limdS}


In the limit $L\rightarrow \infty $, keeping only terms to first order
in $\left( \bar{t}/L\right) $,
\begin{equation*}
{\textstyle{\sqrt{1-\frac{\bar{r}^{2}}{L^{2}}}}}={\textstyle{\ 1-\frac{%
\bar{r}^{2}}{L^{2}}+O\left( \frac{\bar{r}^{4}}{L^{4}}\right) }}~;\; \sinh {%
\textstyle{\ \frac{\bar{t}}{L}=\frac{\bar{t}}{L}+O\left( \frac{\bar{t}^{3}}{%
L^{3}}\right) }}~; \cosh {\textstyle{\frac{\bar{t}}{L}=1+\frac{1}{2}\frac{%
\bar{t}^{2}}{L^{2}}+O\left( \frac{\bar{t}^{4}}{L^{4}}\right) }}~.
\notag
\end{equation*}
The limits of generators $P_{i}$\ and $B_{i}$\ in (\ref{geradores
dS}) can be directly obtained:
\begin{equation}
P_{i}\longrightarrow -{\textstyle{\frac{1}{L}\frac{\bar{t}}{L}}}~\bar{x}%
^{i}~\partial _{\bar{t}}-\bar{\partial}_{i}\,\,\therefore
\,\,P_{i}\longrightarrow -~\bar{\partial}_{i}~,  \label{Pi Minkowski}
\end{equation}%
which we recognize as the \textit{spatial translations} generator on
Minkowski space. On the other hand,
\begin{equation}
B_{i}\longrightarrow \left( \bar{x}^{i}~\partial _{\bar{t}}+\bar{t}~\bar{%
\partial}_{i}\right)  \label{Bi}
\end{equation}%
are the \textit{boost} generators of the Poincar\'{e} algebra. $T_{0}$ and $%
\bar{J}_{i}$ are not affected by the limit $L\rightarrow \infty $, and are
immediately recognized as generators of \textit{\ temporal translations} and
of \textit{rotations}.

For $L\rightarrow \infty $, the dS algebra (\ref{algebra dS}) reduces to:
\begin{equation}
\begin{array}{lll}
\left[ T_{0},B_{i}\right] =-~P_{i}\, & \left[ T_{0},P_{i}\right] =0\, &
\left[ T_{0},\bar{J}_{i}\right] =0\, \\
\left[ B_{i},B_{j}\right] =\epsilon _{ij}^{\;\;k}\bar{J}_{k}\,\quad \quad  &
\left[ B_{i},P_{j}\right] =-~\delta _{ij}~T_{0}\,\quad \quad  & \left[ B_{i},%
\bar{J}_{j}\right] =-~\epsilon _{ij}^{\;\;k}B_{k}\, \\
\left[ P_{i},P_{j}\right] =0\, & \left[ P_{i},\bar{J}_{j}\right] =-~\epsilon
_{ij}^{\;\;k}T_{k}\, & \left[ \bar{J}_{i},\bar{J}_{j}\right] =\epsilon
_{ij}^{\;\;k}\bar{J}_{k}~,%
\end{array}
\label{algebra Poincare}
\end{equation}%
which is exactly the commutation table for the Poincar\'{e} Lie algebra~\cite%
{Ba68}.


\subsection{Limit $L\rightarrow \infty $\ in comoving coordinates}
\label{sec-limFRW}


Let us consider separately the three values of $\kappa$:
\begin{enumerate}
\item $\mathbf{\kappa =0}$: Equations (\ref{generators FRW}) become:
\begin{align}
& S_{0} = \left( \partial _{t}-H~D\right) \,;  \quad S_{i} =
x^{i}~\partial _{t}-{\textstyle{\ \left[ \frac{1}{2}\left( \frac{1}{
a^{2}H}-H~r^{2}\right) \partial _{i}+H~x^{i}~D\right] }} \notag  \\
& T_{i} = -~\partial _{i} \, ; \quad J_{i} =\epsilon
_{ij}^{\;\;k}x^{j}\partial _{k}~, \label{generators FRW k=0}
\end{align}
where, according to Eqs. (\ref{a dS}, \ref{H dS}), $a\left( t\right)
=A\,e^{t/L}$ and $H={\textstyle{\frac{1}{L}}}$. When $L\gg 1$, a divergence
problem turns up in one of the terms of $S_{i}$. In fact,
\begin{equation*}
-{\textstyle{\frac{1}{2}\left( \frac{1}{a^{2}H}\right) }}\partial _{i}=-{%
\textstyle{\frac{1}{2}\left( \frac{L}{Ae^{2\frac{t}{L}}}\right) \partial
_{i}\simeq -{\textstyle{\frac{1}{2A}}L\left( 1-\frac{2t}{L}\right) }}%
\partial _{i}={\textstyle{\frac{1}{2A}}L}}T_{i}+\frac{t}{A}\partial _{i}~.
\end{equation*}%
The first term diverges for $L\rightarrow \infty $. This difficulty can be
overcome by redefining the $S_{i}$\ so as to exclude this undesirable term.
This is fair enough: the problematic term  is linear in other
 generators of the set, $T_{i}$, and any  linear combination
of generators is also a generator. Consider the change
\begin{equation}
S_{i}\rightarrow {\textstyle{\left( \frac{A}{L}~S_{i}+\frac{1}{2A}%
~T_{i}\right) }}=P_{i}~.  \label{combinacao Si}
\end{equation}%
This redefinition not only eliminates the divergence, but also has the form
of $P_{i}$ in FRW coordinates -- according to Table 1 -- which propitiates
the interpretation of the combinations (\ref{combinacao Si})\ as spatial
translation generators. Physical interpretation further suggests the
modification
\begin{equation}
T_{i}\rightarrow {\textstyle{\left( -A~S_{i}+\frac{L}{2A}T_{i}\right) }}%
=-~B_{i}~.  \label{combinacao Ti}
\end{equation}%
We already know that $B_{i}$\ has the meaning of boosts in the Poincar\'{e}
limit. There is no problem for $S_{0}=\left( \partial _{t}-\frac{1}{L}%
~D\right) $ in the limit $L\rightarrow \infty $:
\begin{equation}
\lim_{L\rightarrow \infty }S_{0}=\partial _{t}=T_{0}~,
\label{S0 Minkowski k=0}
\end{equation}%
which is interpreted as the time--translation generator. The rotation
comoving generators $J_{i}$\ are not affected by the limit $L\rightarrow
\infty $. Using Eqs.(\ref{combinacao Si}), (\ref{combinacao Ti})
and (\ref{S0 Minkowski k=0}), it is then an easy task to show that the
commutation table (\ref{[S0,Si]}-\ref{[Ji,Jj]}) coincides with the Poincar%
\'{e} algebra (\ref{algebra Poincare}) when $L\rightarrow \infty $.


\item $\mathbf{\kappa =-1}$: Equations (\ref{generators FRW}) are now
\begin{align}
& S_{0} = \sqrt{1+r^{2}}\left( \partial _{t}-H~D\right)  \,;  \quad
S_{i}  = x^{i}~\partial _{t}-H~\left( \partial _{i}+x^{i}~D\right) \notag   \\
& T_{i}  = -~\sqrt{1+r^{2}}~\partial _{i} \,;  \quad J_{i}  =
\epsilon _{ij}^{\;\;k}x^{j}\partial _{k}~, \label{generators FRW
k=-1}
\end{align}
where
\begin{equation}
H\left( t\right) ={\textstyle{\
\frac{1}{L}+\frac{2L}{e^{2\frac{t}{L}}\left(
A+\sqrt{A^{2}+L^{2}}\right) ^{2}-L^{2}}}} \quad \Rightarrow \quad
\lim_{L\rightarrow \infty }H(t) \simeq {\textstyle{\frac{1}{\left(
A+t\right) }}}~. \notag
\end{equation}
Inserted in (\ref{generators FRW k=-1}), the last expresion above  leads to the
 Poincar\'{e} generators in comoving coordinates. Yet, it is difficult to
attribute the usual interpretation (translations, boosts, rotations)
to the generators in this coordinate system. To recover the
conventional meaning, it is necessary to restore the dimension to the
comoving FRW coordinates~---~see Appendix \ref{app-GRT}, Eqs.
(\ref{Minkowski to FRW k=-1}-\ref{T k=-1}).\ The result is
\begin{align}
& S_{0} =\partial _{\bar{t}}=T_{0}  \, ; \quad \quad \quad \quad
\quad \quad \quad \quad \quad
S_{i} = -~\bar{\partial}_{i}=P_{i} \, ; \notag \\
&T_{i} =-\left( \bar{x}^{i}\frac{\partial }{\partial \bar{t}}+\bar{t}\frac{%
\partial }{\partial \bar{x}^{i}}\right) =-~B_{i} \, ; \quad
J_{i} =\epsilon _{ij}^{\;\;k}\bar{x}^{j}\frac{\partial }{\partial \bar{x}%
^{k}}=\bar{J}_{i}~, \label{generators FRW Minkowski}
\end{align}%
which provide direct interpretations as generators: $S_{0}$ for time
dislocation; $S_{i}$ for spatial translations; $T_{i}$ for boosts and $J_{i}$
for rotations.


\item $\mathbf{\kappa =+1}$: This case is quite analogous to that for $%
\kappa =-1$, up to a point: some imaginary generators appear. This
is to be expected: for $\kappa =0$ and $-1$ the space sections are
open non-compact spaces, while for $\kappa =+1$ it is the compact
sphere $S^{3}$. Translations with parameters $\alpha ^{k}$ on that
sphere, for example, are obtained by exponentiating $\alpha
^{k}S_{k}$, and a factor \textquotedblleft
\emph{i}\textquotedblright\ is necessary to make them finite, as
befits the $S^{3}$ compact group of motions $SO(4)$. The results are
summed up in Table 2.
\end{enumerate}


\begin{center}
\begin{tabular}{||c||}
\hline\hline
In\"{o}n\"{u}-Wigner Contraction for dS \\ \hline\hline
\begin{tabular}{|c|}
\hline
static dS coordinates \\ \hline
\begin{tabular}{|c|c|}
\hline
generator & $L\rightarrow \infty $ version \\ \hline
$T_{0}$ & $\partial _{\bar{t}}$ \\ \hline
$P_{i}$ & $-\bar{\partial}_{i}$ \\ \hline
$B_{i}$ & $\left( \bar{x}^{i}\bar{\partial}_{t}+\bar{t}~\bar{\partial}%
_{i}\right) $ \\ \hline
$\bar{J}_{i}$ & $\epsilon _{ij}^{\;\;k}\bar{x}^{j}\bar{\partial}_{k}$ \\
\hline
\end{tabular}
\\ \hline
\end{tabular}
\\ \hline\hline
\begin{tabular}{|c|}
\hline
comoving FRW coordinates \\ \hline
\begin{tabular}{|c|c|}
\hline
generator & $L\rightarrow \infty $ version \\ \hline
$S_{0}$ &
\begin{tabular}{|c|c|c|}
\hline
$\kappa $ & $r$ and $x^{i}$ without dimension & $r$ and $x^{i}$ with
dimension \\ \hline
$0$ & $\partial _{t}$ & $\partial _{\bar{t}}$ \\ \hline
$-1$ & $\sqrt{1+r^{2}}\left( \partial _{t}-\frac{1}{\left( 1+t\right) }%
D\right) $ & $\partial _{\bar{t}}$ \\ \hline
$+1$ & $\sqrt{1-r^{2}}\left( \partial _{t}-\frac{i}{\left( 1+it\right) }%
D\right) $ & $\partial _{\bar{t}}$ \\ \hline
\end{tabular}
\\ \hline
$S_{i}$ &
\begin{tabular}{|c|c|c|}
\hline
$\kappa $ & $r$ and $x^{i}$ without dimension & $r$ and $x^{i}$ with
dimension \\ \hline
$0$ & $S_{i}\rightarrow \left( \frac{A}{L}S_{i}+\frac{1}{2A}%
T_{i}=P_{i}\right) $ & $-\bar{\partial}_{i}$ \\ \hline
$-1$ & $x^{i}\partial _{t}-\frac{1}{\left( 1+t\right) }\left( \partial
_{i}+x^{i}D\right) $ & $-\bar{\partial}_{i}$ \\ \hline
$+1$ & $x^{i}\partial _{t}+\frac{i}{\left( 1+it\right) }\left( \partial
_{i}-x^{i}D\right) $ & $-i\left( -\bar{\partial}_{i}\right) $ \\ \hline
\end{tabular}
\\ \hline
$T_{i}$ &
\begin{tabular}{|c|c|c|}
\hline
$\kappa $ & $r$ and $x^{i}$ without dimension & $r$ and $x^{i}$ with
dimension \\ \hline
$0$ & $T_{i}\rightarrow {{\left( -A~S_{i}+\frac{L}{2A}T_{i}\right) }}=-B_{i}$
& $-\left( \bar{x}^{i}\bar{\partial}_{t}+\bar{t}~\bar{\partial}_{i}\right) $
\\ \hline
$-1$ & $-\sqrt{1+r^{2}}\partial _{i}$ & $-\left( \bar{x}^{i}\bar{\partial}%
_{t}+\bar{t}~\bar{\partial}_{i}\right) $ \\ \hline
$+1$ & $-\sqrt{1-r^{2}}\partial _{i}$ & $-i\left( \bar{x}^{i}\bar{\partial}%
_{t}+\bar{t}~\bar{\partial}_{i}\right) $ \\ \hline
\end{tabular}
\\ \hline
$J_{i}$ &
\begin{tabular}{|c|c|}
\hline
$r$ and $x^{i}$ without dimension & $r$ and $x^{i}$ with dimension \\ \hline
$\epsilon _{ij}^{\;\;k}x^{j}\partial _{k}$ & $\epsilon _{ij}^{\;\;k}\bar{x}%
^{j}\bar{\partial}_{k}$ \\ \hline
\end{tabular}
\\ \hline
\end{tabular}
\\ \hline
\end{tabular}
\\ \hline\hline
\end{tabular}

\bigskip

\emph{Table 2: Comparison between  contractions of the dS generators in
the static and comoving systems.}

\end{center}
Generators $S_{0}$, $S_{i}$, $T_{i}$, and $J_{i}$\ are suitable for
recognizing the symmetry breaking between dS and FRW: $S_{0}$ and $S_{i}$
disappear unless $a\left( t\right) $ is a dS solution. They are not, however, convenient
for comparison with the Poincar\'{e}  generators when $%
L\rightarrow \infty $: length dimension must be restored to space
coordinates $r$, $x^{i}$ beforehand. Once this is done the $T_{i}$, for example, which
in comoving coordinates might be associated to spatial translations (in
spatial sections with $\kappa =-1,0,+1$), acquire the meaning of boosts when
written in barred coordinates (with dimension). And the problem raised at the end of Section \ref%
{sec-KillFRW} has  a simple answer: the de Sitter generators $\{S_{k}\}$ are not to be regarded, by themselves,  as spanning the space section. Only some linear combinations
of the $S_k$ and the $T_k$ are. There is consequently no problem in
considering a possible passage from a $dS$ model to a Standard Model.




\section{Final Remarks \label{sec-FinalRemmarks}}


The FRW coordinates are suitable to examine the relationship between the $dS$
and Standard Model symmetries, but the static coordinates are more
appropriate to obtain physical interpretations of their generators. For the
nowadays observationally favored $\kappa = 0$ case, boosts and translations
are certain linear combinations of the FRW Killing fields, with factors
necessary to adapt their dimensions.

It goes without saying that many questions remain in the open. For instance,
how could it happen that matter/radiation be ``added'' to a primeval $dS$
Universe? We have, quite naturally, supposed the cosmological \emph{constant}
$\Lambda$ to be really \emph{constant}\, in coordinate time. There is no
contradiction neither with Einstein's equations nor with energy conservation
to have a time-dependent $\Lambda(t)$~---~provided mater/radiation is
created or destroyed so as to maintain the energy balance~\cite{ABP05}. An
evolving $\Lambda(t)$, with very high values~---~providing for
inflation~---~at the primordial Universe but subsequently (and quickly)
decreasing towards it present stable value, could provide a model. The
negative point is, of course, the absence of any dynamics for $\Lambda(t)$%
~---~which, by the way, is equivalent to a scalar field in a homogeneous
Universe. Such a dynamics is usually put by hand, as in the inflaton
scenery. It would be preferable that such dynamics be brought forth by some
fundamental theory or principle.

The problem of the transition between accelerated and decelerated
cosmological solutions is certainly more embracing than that of the
transition between the dS and FRW paradigms. A suitable modeling of the
equation of state for the cosmic fluid can describe the transition -- an
example is given in \cite{Dymni86}, and Ref.\cite{Al05(3)} is a effort in
that direction.

It is a fascinating point that the Levi-Civita connection for dS satisfies a
Yang-Mills equation, so that $dS$ spacetimes are, in a sense, also gauge
theories. A gauge theory has its proper dynamics. Partial breaking of gauge
theories are no novelty, but here it would involve spacetime itself. An
attempt, using the formalism of extended gauge theories \cite{Al95(2)}, is
under study.


\section*{\protect\small Acknowledgements}

An unknown referee is thanked for useful suggestions. R.A. thanks to
Conselho Nacional de Pesquisas (CNPq), Brazil and R.R.C. and L.G.M.
are grateful to Funda\c{c}\~{a}o de Amparo \`{a} Pesquisa do Estado
de S\~{a}o Paulo (FAPESP), Brazil, for support (grants 02/05763-8
and 02/10263-4 respectively).


\appendix


\section{The Killing components $\protect\xi _{\protect\mu }$}
\label{app-Killcomponents}

We obtain here the component $\xi_0$ when it does not vanish (that is, in
the dS case $U=\kappa$), as a typical example of the general integration
procedure. The steps consist mainly of taking crossed derivatives of Eqs.(%
\ref{I}-\ref{X}), up to a point in which a decoupled equation for the $%
x^{\nu }$-dependence of the component $\xi _{\mu }$ is obtained.

The radial dependence of $\xi _{0}$\ is given by the integration of
(\ref{diff eq xi0 r}). Substituting $U=\kappa $, we rewrite it as
\begin{equation*}
\partial _{1}\left[ \left( 1-\kappa r^{2}\right) ~\partial _{1}\xi
_{0}+\kappa r~\xi _{0}\right] =0~.
\end{equation*}
The term between square-brackets must be a constant in $r$, but not with
respect to the others coordinates:
\begin{equation}
\left( 1-\kappa r^{2}\right) ~\partial _{1}\xi _{0}+\kappa r~\xi _{0}=\tau
_{1}\left( \theta ,\phi \right) ~,  \label{diff eq xi0 Tau1}
\end{equation}
whose solution is the homogeneous  ($\tau _{1}=0$) solution plus a
particular one, $\xi _{0}=\xi _{0}^{h}+\xi _{0}^{p}$. The first
comes immediately: $\xi _{0}^{h}=\tau _{2}\left( \theta ,\phi
\right) \sqrt{1-\kappa r^{2}}$. A particular solution of (\ref{diff
eq xi0 Tau1}) is of the form $\xi _{0}^{p}=\tau _{1}\left( \theta
,\phi \right)
r$, as easily verified by substitution. Thus,%
\begin{equation}
\xi _{0}=\tau _{1}\left( \theta ,\phi \right) r+\tau _{2}\left( \theta ,\phi
\right) \sqrt{1-\kappa r^{2}}~,  \label{xi0 r-dependence}
\end{equation}
where $\tau _{1}\left( \theta ,\phi \right) $\ and $\tau _{2}\left( \theta
,\phi \right) $\ must be obtained from the other equations of the system (%
\ref{I}-\ref{X}).

The $\theta $-dependence of $\xi _{0}$ is obtained by taking $\partial _{0}$
of Eq.(\ref{IX}), using (\ref{III}), (\ref{II}) and (\ref{I}) to eliminate
respectively $\partial _{0}\xi _{2}$, $\partial _{0}\xi _{1}$\ and $\partial
_{0}\xi _{0}$\ and then using Eq.(\ref{IX}) to substitute $\partial _{2}\xi
_{2}$. The result is
\begin{equation}
\partial _{2}\partial _{2}\xi _{0}+r\left( 1-\kappa r^{2}\right) \partial
_{1}\xi _{0}+U~r^{2}~\xi _{0}=0.  \label{diff eq xi0 theta}
\end{equation}
From (\ref{xi0 r-dependence}) follows
\begin{equation*}
\left[ \partial _{2}\partial _{2}\tau _{1}\left( \theta ,\phi \right) +\tau
_{1}\left( \theta ,\phi \right) \right] ~r+\partial _{2}\partial _{2}\tau
_{2}\left( \theta ,\phi \right) ~\sqrt{1-\kappa r^{2}}=0~.
\end{equation*}
If the above equality is to be verified for any value of $r$, the
coefficients of the functions $r$ and $\sqrt{1-\kappa r^{2}}$ must vanish
independently:
\begin{equation}
\partial _{2}\partial _{2}\tau _{1}\left( \theta ,\phi \right) +\tau
_{1}\left( \theta ,\phi \right) =0~,\text{ and} \; \quad \partial
_{2}\partial _{2}\tau _{2}\left( \theta ,\phi \right) =0~.
\label{system of a Tau theta}
\end{equation}
from which we have
\begin{equation}
\tau _{1}\left( \theta ,\phi \right) =\tau _{3}\left( \phi \right)
\sin \theta +\tau _{4}\left( \phi \right) \cos \theta ~,\text{ and}
\; \quad \tau _{2}\left( \theta ,\phi \right) =\tau _{5}\left( \phi
\right) ~\theta +\tau _{6}\left( \phi \right) ~.  \label{Tau1 and
Tau2}
\end{equation}
This concludes the $\theta $-dependence's determination for $\xi _{0}$:
inserting the last two expressions in (\ref{xi0 r-dependence}),
\begin{equation}
\xi _{0}=r~\left[ \tau _{3}\left( \phi \right) \sin \theta +\tau _{4}\left(
\phi \right) \cos \theta \right] +\sqrt{1-\kappa r^{2}}\left[ \tau
_{5}\left( \phi \right) ~\theta +\tau _{6}\left( \phi \right) \right] ~.
\label{xi0 theta-dependence}
\end{equation}

Concerning the $\phi $-dependence: calculate $\partial _{0}$ of (\ref{X})
and use (\ref{IV}), (\ref{III}), (\ref{II}) and (\ref{I}) to eliminate $%
\partial _{0}\xi _{3}$, $\partial _{0}\xi _{2}$, $\partial _{0}\xi _{1}$\
and $\partial _{0}\xi _{0}$\ respectively. Then, substitute $\partial
_{3}\xi _{3}$ as given by (\ref{X}), to find
\begin{equation}
\partial _{3}\partial _{3}\xi _{0}+\sin \theta \cos \theta ~\partial _{2}\xi
_{0}+r\left( 1-\kappa r^{2}\right) \sin ^{2}\theta ~\partial _{1}\xi
_{0}+U~r^{2}\sin ^{2}\theta ~\xi _{0}=0~.  \label{diff eq xi0 phi}
\end{equation}
Again recalling that $U=\kappa $ in the present case, insertion of (\ref{xi0
theta-dependence})\ in (\ref{diff eq xi0 phi}) yields
\begin{align}
& r\sin \theta \left[ \partial _{3}\partial _{3}\tau _{3}\left( \phi
\right) +\tau _{3}\left( \phi \right) \right] +r\cos \theta \left[
\partial_{3}\partial _{3}\tau _{4}\left( \phi \right) \right] \notag \\
& +\sqrt{1-\kappa r^{2}} \left[ \partial _{3}\partial _{3}\tau
_{5}\left( \phi \right) ~\theta +\sin \theta \cos \theta ~\tau
_{5}\left( \phi \right) +\partial _{3}\partial _{3}\tau _{6}\left(
\phi \right) \right] =0~. \notag
\end{align}
In view of the independence of the coefficients $r\sin \theta $, $r\cos
\theta $\ and $\sqrt{1-\kappa r^{2}}$, we arrive at the system
\begin{equation}
\partial _{3}\partial _{3}\tau _{3}\left( \phi \right) +\tau _{3}\left( \phi
\right) =0~,\; \quad \partial _{3}\partial _{3}\tau _{4}\left( \phi
\right) =0~, \text{ and} \; \quad \partial _{3}\partial _{3}\tau
_{5}\left( \phi \right) ~\theta +\sin \theta \cos \theta ~\tau
_{5}\left( \phi \right) +\partial _{3}\partial _{3}\tau _{6}\left(
\phi \right) =0~. \label{system of a Tau phi}
\end{equation}%
From the first two equations in (\ref{system of a Tau phi}),
\begin{equation}
\tau _{3}\left( \phi \right) =\tau _{7}\sin \phi +\tau _{8}\cos \phi
~,\text{ and} \; \quad \tau _{4}\left( \phi \right) =\tau _{9}~\phi
+\tau _{10}~. \label{Tau3 and Tau4}
\end{equation}%
The third equation imposes
\begin{equation}
\tau _{5}\left( \phi \right) =0~,\text{ and} \; \quad \partial
_{3}\partial _{3}\tau _{6}\left( \phi \right) =0~,  \label{Tau5}
\end{equation}%
thanks to the independence of the functions that appear multiplying $%
\partial _{3}\partial _{3}\tau _{5}\left( \phi \right) $, $\tau _{5}\left(
\phi \right) $\ and $\partial _{3}\partial _{3}\tau _{6}\left( \phi \right) $%
. From the last result,
\begin{equation}
\tau _{6}\left( \phi \right) =\tau _{11}~\phi +\tau _{12}~.  \label{Tau6}
\end{equation}%
Substituting (\ref{Tau3 and Tau4}-\ref{Tau6}) in (\ref{xi0
theta-dependence}) we determine the $\phi $-dependence of $\xi
_{0}$:
\begin{equation}
\xi _{0}=r~\left[ \left( \tau _{7}\sin \phi +\tau _{8}\cos \phi \right) \sin
\theta +\left( \tau _{9}~\phi +\tau _{10}\right) \cos \theta \right] +\sqrt{%
1-\kappa r^{2}}\left[ \tau _{11}~\phi +\tau _{12}\right] ~.
\label{xi0 phi-dependence}
\end{equation}%
The periodicity condition $\xi _{0}\left( r,\theta ,0\right) =\xi _{0}\left(
r,\theta ,2\pi \right) $ on the azimuthal coordinate gives $\tau _{9}=\tau
_{11}=0$ which, in (\ref{xi0 phi-dependence}), leads to
\begin{equation*}
\xi _{0}=r~\left[ \left( \tau _{7}\sin \phi +\tau _{8}\cos \phi \right) \sin
\theta +\tau _{10}\cos \theta \right] +\sqrt{1-\kappa r^{2}}~\tau _{12}~.
\end{equation*}%
Renaming the constants $\tau _{7}=C_{2}; \; \tau _{8}=C_{1}; \; \tau
_{10}=C_{3}; \; \tau _{12}=C_{0}$, we arrive at the final form:%
\begin{equation}
\xi _{0}=C_{0}\sqrt{1-\kappa r^{2}}+r~\left[ \left( C_{2}\sin \phi
+C_{1}\cos \phi \right) \sin \theta +C_{3}\cos \theta \right] ~;
\label{xi0 cov}
\end{equation}

The other covariant Killing components are found by analogous operations and
are:
\begin{align}
\xi _{1}=& \; a^{2}H \frac{r}{\sqrt{1-\kappa
r^{2}}}~C_{0}+{\textstyle{\left[ \frac{\left( 1-\kappa ^{2}\right)
}{2}\left( \frac{1}{H}+a^{2}H~r^{2}\right) -\kappa ~a^{2}H\right]
}}\left[ \left( C_{2}\sin \phi +C_{1}\cos \phi
\right) \sin \theta +C_{3}\cos \theta \right]  \notag \\
&+~a^{2}~\frac{1}{\sqrt{1-\kappa r^{2}}}\left[ \left( K_{1}\cos \phi
+K_{2}\sin \phi \right) \sin \theta +K_{3}\cos \theta \right] ~;
\label{xi1 cov} \\
\xi _{2}=& \, {\textstyle{\left[ \frac{\left( 1-\kappa ^{2}\right)
}{2}\left( \frac{1}{H}-a^{2}Hr^{2}\right) -\kappa ~a^{2}H\right]
}}r\left[ \left( C_{2}\sin \phi +C_{1}\cos \phi \right) \cos \theta
-C_{3}\sin \theta \right]
\notag \\
&+~a^{2}~r\sqrt{1-\kappa r^{2}}~\left[ \left( K_{1}\cos \phi
+K_{2}\sin \phi \right) \cos \theta -K_{3}\sin \theta \right]
+~a^{2}~r^{2}~\left( L_{1}\sin \phi -L_{2}\cos \phi \right) ~;
\label{xi2 cov} \\
\xi _{3}=& \, {\textstyle{\left[ \frac{\left( 1-\kappa ^{2}\right)
}{2}\left( \frac{1}{H}-a^{2}Hr^{2}\right) -\kappa ~a^{2}H\right]
}}r\left( C_{2}\cos \phi -C_{1}\sin \phi \right) \sin \theta  \notag \\
& +a^{2}r\sqrt{1-\kappa r^{2}}\left(
K_{2}\cos \phi -K_{1}\sin \phi \right) \sin \theta  \notag \\
&+~a^{2}~r^{2}~\left[ \sin \theta \cos \theta ~\left( L_{1}\cos \phi
+~L_{2}\sin \phi \right) -\sin ^{2}\theta ~L_{3}\right] ~.
\label{xi3 cov}
\end{align}
Each Killing vector field $\xi ^{(P)}$ will be obtained by choosing a
convenient, simple value (such as $\pm 1$) for one of the constants while
putting all the other $=0$. In particular, that we have complete freedom in
the choice of overall signs. Of course, linear combinations of such fields
will also be solutions. Case ${\dot{U}}=0$ contributes the whole of Eq.(\ref%
{xi0 cov}) and the terms in $C_{\alpha }$ of Eqs.(\ref{xi1 cov}, \ref{xi2
cov}, \ref{xi3 cov}). Indeed, case ${\dot{U}}\neq 0$ corresponds to $%
C_{0}=C_{1}=C_{2}=C_{3}=0$, vindicating the statement of Section \ref%
{sec-KillFRW}: four of the Killing vector fields vanish completely.

Let us go back to the FRW Killing fields (\ref{Killing fields}). Two steps
are necessary to arrive at their most convenient form: \textbf{(i)} Find the
contravariant components, by applying to the components (\ref{xi0 cov}---\ref%
{xi3 cov}) the inverse metric
\begin{equation*}
(g^{\mu \nu })={\mbox{diag}}~\left( 1,-\,\frac{(1-\kappa r^{2})}{a^{2}},-\,%
\frac{1}{a^{2}r^{2}},-\,\frac{1}{a^{2}r^{2}\sin ^{2}\theta }\right) \,;
\end{equation*}%
and \textbf{(ii)} Simpler, more recognizable expressions obtain if we change
from the above basis $\{\partial _{t},\partial _{r},\partial _{\theta
},\partial _{\phi }\}$ to basis $\{\partial _{x}$, $\partial _{y}$, $%
\partial _{z}\}$, related to cartesian variables on the space--section, $%
x=r\sin \theta \cos \phi \;y=r\sin \theta \sin \phi \;;z=r\cos \theta $. It
will be also convenient to introduce the space--section dilation generator $%
D=r\partial _{r}=x\,\partial _{x}+y\,\partial _{y}+z\,\partial _{z}$. After
this change, go back to notation $x^{1}=x$, $x^{2}=y$, $x^{3}=z$. The result
are the operators (\ref{generators FRW}).


\section{The generalized Robertson transformation}

\label{app-GRT}

The simultaneous use of de Sitter static line element (\ref{ds2 dS})
and the standard form (\ref{ds2 FRW}) of the FRW interval has led to
very interesting results~\cite{Dymni86,Gliner70}. We give now
explicit expressions for the transformations relating them.

The simple Robertson transformations \cite{To87,Wei72}
\begin{eqnarray}
\bar{r} &=&r\left( Ae^{t/L}\right) ~,  \label{r bar Robertson} \\
\bar{t} &=&t-{\textstyle{\frac{L}{2}}}\,\ln {\textstyle{\left[ 1-\frac{r^{2}%
}{L^{2}}\left( Ae^{t/L}\right) ^{2}\right] }}~,  \label{t bar Robertson}
\end{eqnarray}%
convert the de Sitter static line element into the form
\begin{equation}
ds^{2}=dt^{2}-\left( Ae^{t/L}\right) \left[ dr^{2}+r^{2}d\theta
^{2}+r^{2}\sin ^{2}\theta d\phi ^{2}\right] ~,  \label{ds2 Robertson}
\end{equation}%
exactly the FRW interval for the particular value $\kappa =0$ with $%
a(t)=Ae^{ct/L}$, the paradigmatic inflationary solution. Our aim is to
extend these transformations so as to include cases $\kappa =\pm 1$. We look
for a transformation
\begin{eqnarray}
\bar{r} &=&rf(t)~,  \notag \\
\bar{t} &=&g(t)+h(t,r)~,  \label{transf guess}
\end{eqnarray}%
in terms of functions $f(t)$, $g(t)$ and $h(t,r)$ to be determined.
Inserting these expressions in the static interval (\ref{ds2 dS}) and
imposing equality with the FRW interval leads to a system of partial
differential equations,
\begin{gather}
{\textstyle{\ \left[ 1-\frac{\left( rf\right) ^{2}}{L^{2}}\right] }}\left(
\dot{g}+\dot{h}\right) ^{2}-{\textstyle{\frac{1}{\left[ 1-\frac{\left(
rf\right) ^{2}}{L^{2}}\right] }}}r^{2}\dot{f}^{2}=1~,  \label{one} \\
{\textstyle{\left[ 1-\frac{\left( rf\right) ^{2}}{L^{2}}\right] }}2h^{\prime
}\left( \dot{g}+\dot{h}\right) -{\textstyle{\frac{1}{\left[ 1-\frac{\left(
rf\right) ^{2}}{L^{2}}\right] }}}2rf\dot{f}=0~,  \label{two} \\
{\textstyle{\frac{1}{\left[ 1-\frac{\left( rf\right) ^{2}}{L^{2}}\right] }}}%
f^{2}-{\textstyle{\left[ 1-\frac{\left( rf\right) ^{2}}{L^{2}}\right] }}%
h^{\prime 2}=\frac{f^{2}(t)}{1-\kappa r^{2}}~,  \label{three} \\
f^{2}(t)=a^{2}(t)~,  \label{four}
\end{gather}%
where the dots and primes indicate derivations with respect to $t$\ and $r$,
respectively. Eliminating the factor $(\dot{g}+\dot{h})$ from the first two
equations, it is possible to determine $h^{\prime 2}$ in terms of $f$ and $%
\dot{f}$. Substituting the result into Eq.(\ref{three}) leads to
\begin{equation}
\dot{f}^{2}=\frac{1}{L^{2}}f^{2}-\kappa ,  \label{diff eq f}
\end{equation}%
which is, in view of Eq.(\ref{four}), just the Friedmann equation ${\dot{a}}%
^{2}={\textstyle}\frac{1}{L^{2}}a^{2}-\kappa $ for the de Sitter case with $%
\kappa =0,\pm 1$ -- see Ref. \cite{ACM05}, for example. This is a curious
fact, since the Friedmann equations come from the gravitational field
dynamics~---~dictated by Einstein's equations~---~and the FRW interval,
while relation (\ref{diff eq f}) rises solely as a coordinate transformation
effect. Its solution is, of course, just Eq.(\ref{a dS}):
\begin{equation}
f(t)=A\cosh \frac{t}{{L}}+\sqrt{A^{2}-\kappa L^{2}}\sinh \frac{t}{{L}}~.
\label{f}
\end{equation}%
The next step is to isolate $h^{\prime }$ in Eq.(\ref{three}), and integrate
directly, thereby fixing the radial dependence of $h(t,r)$. It is safer to
solve for each value of $\kappa $ separately and afterwards gather the
results in a single formula, which is
\begin{equation}
h(t,r)=L~\ln {\textstyle{\ \left[ \frac{\sqrt{\left( \frac{f}{L}\right)
^{2}-\kappa }+\frac{f}{L}\sqrt{1-\kappa r^{2}}}{\sqrt{1-\frac{(rf)^{2}}{L^{2}%
}}}\right] }}+z(t)~.  \label{h}
\end{equation}%
For the time being, $z(t)$ is arbitrary and $\bar{t}$\thinspace\ is
determined up to a function of $t$, say, $\bar{g}(t)\equiv g(t)+z(t)$. The
function $\bar{g}(t)$ will be fixed as soon as we evaluate $\dot{h}(t,r)$.
The derivatives of Eq.(\ref{h}) with respect to $t$ and $r$, once inserted
into Eq.(\ref{two}), give:
\begin{equation}
\frac{d\bar{g}}{dt}=0,\quad {\mbox{i.e.,}}\quad \bar{g}=g+z={\mbox{constant}}%
\equiv T~.  \label{g bar}
\end{equation}%
Using Eqs.(\ref{f}), (\ref{h}) and (\ref{g bar}), Eqs. (\ref{transf guess})
can be rewritten as
\begin{eqnarray}
\bar{r}(t,r) &=&ra(t)~;  \label{r bar RGT} \\
\bar{t}(t,r) &=&T+L\ln {\textstyle{\ \left[ \frac{\sqrt{\left[ \frac{a(t)}{L}%
\right] ^{2}-\kappa }+\frac{a(t)}{L}\sqrt{1-\kappa r^{2}}}{\sqrt{1-\frac{%
[ra(t)]^{2}}{L^{2}}}}\right] }}~.  \label{t bar RGT}
\end{eqnarray}

Constant $T$ can be fixed by requiring that these expressions remain
non-singular in the limit $L\rightarrow \infty $. This strong requirement
comes from two facts: \textit{(i)} we will study the In\"{o}n\"{u} -Wigner
contraction of the dS symmetry operators into Poincar\'{e} generators in
both static and comoving coordinates, comparing the results afterwards via (%
\ref{r bar RGT},\ref{t bar RGT}), which, therefore, must hold; \textit{(ii)}
the dS static line element (\ref{ds2 dS}) reduces mildly to Minkowski
interval in spherical coordinates,
\begin{equation}
ds^{2}=d\bar{t}^{2}-d\bar{r}^{2}-\bar{r}^{2}\left( d\theta ^{2}+\sin
^{2}\theta ~d\phi ^{2}\right) ,  \label{ds2 Minkowski}
\end{equation}%
in limit $L\rightarrow \infty $. If we substitute the dS scale factor (\ref%
{a dS})\ in FRW line element (\ref{ds2 FRW}), what we obtain is the dS
interval in comoving coordinates. This form of $ds^{2}$\ must also reduce to
Minkoski interval (\ref{ds2 Minkowski}) after taking $L\rightarrow \infty $
and restoring the dimension of length to the (dimensionless) radial comoving
coordinate $r$. To restore the dimension of $r$ means to convert $(t,r)$
into $\left( \bar{t},\bar{r}\right) $ through (\ref{r bar RGT},\ref{t bar
RGT}) with $L\rightarrow \infty $.

Let us proceed case by case.

\begin{itemize}
\item For $\kappa =0$, $a(t)=Ae^{t/L}$ and (\ref{t bar RGT}) is:%
\begin{equation}
\bar{t}=T+L\ln {\textstyle{\frac{2A}{L}}}+t - {\textstyle{\frac{1}{2}}} L\ln
{\textstyle{\ \left[ 1-\frac{r^{2}}{L^{2}} \left( Ae^{t/L}\right) ^{2}\right]
}} \;\;\;\left( \kappa =0\right) ~.  \label{t bar k=0}
\end{equation}%
Notice that in the limit $L\rightarrow \infty $\ the second term diverges.
In order to avoid this singularity, we fix
\begin{equation}
T=-\, L\ln {\textstyle{\frac{2A}{L}}} \;\;\;\left( \kappa =0\right) ~.
\label{T k=0}
\end{equation}%
This choice is specially convenient, as (\ref{t bar k=0}) turns to be
exactly the traditional Robertson transformation law (\ref{t bar Robertson}).


\item In the case $\kappa =-1$, if $L\gg 1$,
\begin{eqnarray}
\bar{r} &=&ra\left( t\right) ~,\;\;\;a\left( t\right) =\left( A+t\right) ~,
\notag \\
\bar{t} &=&T+a\left( t\right) \sqrt{1+r^{2}}\;\;\;\;\;\;\;\left( L\gg
1~,\kappa =-1\right) ~.  \label{Minkowski to FRW k=-1}
\end{eqnarray}%
These are exactly the expressions for the coordinate transformation law from
Minkowski (barred) coordinates to FRW coordinates. This statement is
verified, for example, by repeating the procedures of this section for line
elements (\ref{ds2 Minkowski}) and (\ref{ds2 FRW}): start with an anzatz
just like (\ref{transf guess}), then substitute it in (\ref{ds2 Minkowski})
to find, after comparison with (\ref{ds2 FRW}), equations analogous to (\ref%
{one}-\ref{four}), which, by direct integration, lead to (\ref{Minkowski to
FRW k=-1}). This leads to the conclusion that we may take
\begin{equation}
T=0\;\;\;\left( \kappa =-1\right)  \label{T k=-1}
\end{equation}%
without loss of generality. Of course, one might prefer to take another
value for $T$ in this case, e.g., $T=-L\ln \left( 1+\frac{2A}{L}\right) $.
The only constraint is to avoid functional forms of $T$ that diverge in the
limit $L\rightarrow \infty $.


\item If $\kappa =+1$ and $L\gg 1$, Eqs.(\ref{r bar RGT},\ref{t bar RGT})
reduce to:
\begin{eqnarray}
\bar{r} &=&ra\left( t\right) ~,\;\;\;a\left( t\right) =\left( A+i~t\right) ~,
\notag \\
\bar{t} &=&T+L\ln i-i~a\left( t\right) \sqrt{1-\bar{r}^{2}}\;\;\;\left( L\gg
1~,\kappa =+1\right) ~.  \label{Minkowski to FRW k=+1}
\end{eqnarray}%
The divergence of the second term is prevented by the choice
\begin{equation}
T=-\,L\,\ln i\text{ }\;\;\;\left( \kappa =+1\right) ~.  \label{T k=+1}
\end{equation}

The complex factors turning up are artifacts of the coordinate system (see
the comments below Eq.(\ref{generators FRW Minkowski}), item 3). They
reflect our forceful requirement of equality between the FRW line element
\begin{equation}
ds^{2}=dt^{2}-a^{2}\left( t\right) \left[ \frac{dr^{2}}{1-r^{2}}+r^{2}\left(
d\theta ^{2}+\sin ^{2}\theta d\phi ^{2}\right) \right] ~.
\label{ds2 FRW k=+1}
\end{equation}
and the Minkowski interval. Minkowski space is flat, non-compact, while the
space section of the line element above is a sphere.
\end{itemize}

In conclusion, we use the definitions
\begin{equation}
T=L\,\ln \tau \left( \kappa \right) ~;\;\;\;\tau \left( \kappa \right)
=\left\{
\begin{array}{c}
\frac{L}{2A}~,\;\;\text{\ }\kappa =0 \\
1~,\;\;\;\kappa =-1 \\
\frac{1}{i}~,\;\;\;\kappa =+1%
\end{array}%
\right.  \label{T(K)}
\end{equation}%
and the identity (obvious for the dS solution)
\begin{equation}
{\textstyle{\ \sqrt{\left[ \frac{a\left( t\right) }{L}\right] ^{2}-\kappa }}}%
=a\left( t\right) H\left( t\right)  \label{a H}
\end{equation}%
to write the final unified form of the generalized transformation as Eq. (%
\ref{TRG}) of the main text.

Notice that what was done in this section differs from the results found in
the literature. It does not coincide with the treatments presented, for
example, in \cite{Ha95}, \cite{Lo74}, \cite{Ki02} and \cite{Mu05}, but
implements in all details the transformation mentioned in \cite{Gliner70}.
Reference \cite{Ha95} exhibits interval (\ref{ds2 FRW k=+1}) -- up to a
signature convention and the definition $r=\sin \chi $. However, the
expansion parameter it is taken to be $a\left( t\right) =L\cosh \frac{t}{L}$%
, which would correspond to arbitrarily fixing $A=L$ in our solution (\ref{a
dS}). This would imply that the scale factor is always greater than the dS
horizon $L$. Besides, no transformation between dS static coordinates and
comoving ones is given, but only transformations from embedding
(5-dimensional) variables to comoving coordinates. Choice $A=L$ is made also
in ref. \cite{Er95},\ which presents an extensive study of dS geometry in
various coordinate systems, including the transformation between the static
and the three FRW sections of dS space-time. Nevertheless, those are not
exactly the generalized Robertson transformations; not only because of the
choice for the value of $A$, but also in view of the fact that they do not
have an appropriate limit when $L\rightarrow \infty $.


\end{document}